\documentclass[11pt,showpacs,nofootinbib]{revtex4-1}

\usepackage{epsfig}
\input epsf.tex
\usepackage{amssymb}
\usepackage{amsmath}
\usepackage{amsthm}
\usepackage{amsopn}

\def\comment#1{}

\def\beq{\begin{equation}}
\def\eeq{\end{equation}}
\def\bea{\begin{eqnarray}}
\def\eea{\end{eqnarray}}

\usepackage{ulem}
\usepackage{cancel}
\usepackage{color}

\begin{document}
	{\textsf{\today}
		\vspace*{1cm}
		\title{ B-mode Power Spectrum of CMB via  Polarized Compton Scattering }
		\author{  Jafar Khodagholizadeh$^{1}$\footnote{j.gholizadeh@cfu.ac.ir}, Rohoollah Mohammadi$^{2,3}$\footnote{rmohammadi@ipm.ir}, Mahdi Sadegh$^{4}$\footnote{m.sadegh@ipm.ir}and Ali Vahedi $^{5}$\footnote{vahedi@khu.ac.ir,vahedi@ipm.ir}}
		\affiliation{$^{1}$ Farhangian University, P.O. Box 11876-13311,  Tehran, Iran.\\
			$^{2}$Iranian National Museum of Science and Technology (INMOST), PO BOX: 11369-14611, Tehran, Iran. \\
			$^{3}$School of Astronomy, Institute for Research in Fundamental Sciences (IPM), P. O. Box 19395-5531, Tehran, Iran.\\
			$^{4}$ School of Particles and Accelerators, Institute for Research in Fundamental Sciences (IPM), P. O. Box 19395-5531, Tehran, Iran.\\
			$^{5}$ Department of Physics, Kharazmi University, Shahid Mofateh Ave, Tehran, Iran.}
		
		\begin{abstract}
			In this work, some evidences for existing an asymmetry in the density of left- and right-handed cosmic electrons ($\delta_L$ and $\delta_R$ respectively) in universe motivated us to calculate the dominated contribution of this asymmetry in the generation of the B-mode power spectrum. In the standard cosmological scenario, Compton scattering in the presence of scalar matter perturbation can not generate magnetic-like pattern in linear polarization  $C_{ B\,l}^{(S)}=0$ while in the case of polarized Compton scattering $C_{B\,l}^{(S)}\propto \delta_L^2$. By adding up the power spectrum of the B-mode generated by the polarized Compton scattering to power spectra produced by weak lensing effects and Compton scattering in the presence of tensor perturbations, we show that there is a significant amplification in $C_{B\,l}$ in large scale $l<500$ for $\delta_L>10^{-6}$, which can be observed in future experiments.  Finally, we have shown that $C_{ B\,l}^{(S)}$ generated by polarized Compton scattering can suppress the tensor-to-scalar ratio ($r$-parameter) so that this contamination can be comparable to a primordial tensor-to-scalar ratio spatially for $\delta_L>10^{-5}$.
		\end{abstract}
		
		
		\maketitle
		\section{Introduction}	
		Primordial gravitational waves (PGWs) indirectly affect the CMB temperature and the polarization (in the context of the standard scenario of Big Bang) in the low-frequency range $(\sim 10^{-18}Hz-10^{-16}Hz)$ \cite{61,71}.
		Also, the primordial gravitational waves can generate a magnetic-like component pattern for linear polarization of CMB, called $B$-modes polarization \cite{Sejak}. The amplitude of this signal is characterized by the tensor-to-scalar ratio ($r$-parameter) at the power spectrum level. One of the most important constraints in this regard comes from combining Bicep/Keck data with Planck and WMAP data, which reports $r<0.07$ at $95\%$ confidence \cite{Ade2018}. In comparison, other experiments such as POLARBEAR \cite{Polar}, BICEP/Keck \cite{Bi, Ke} and SPT \cite{Spt} collaborations try to improve the precision in the $B$-mode power spectrum as well as $r$-parameter. There exist other detectors such as the QUIJOTE and SPIDER. QUIJOTE is an experiment designed to measure $B$-mode polarization. Also it is sufficiently sensitive to detect a primordial gravitational wave amplitude around $r=0.05$ \cite{Rub1, Rub2}. On the other hand, SPIDER is a balloon-borne instrument designed to detect the polarization of the millimeter-wave sky and its goal is to detect the divergence-free mode of primordial gravitational waves in CMB radiation \cite{Spider}. The measurement of the B-mode polarization in the CMB induced by primordial gravitational waves \cite{Marc} can be used to provide an independent cross-check of the early-universe expansion history \cite{Donghui}. Also, independent of Planck observations, the morphology of E and B maps of Galactic dust emission have been explored in \cite{Weil}. In this regard, an augmented version of dual messenger algorithm \cite{Kodi,Kodi1} can be used for the separation of  pure $E/B$ decomposition on the sphere, based on the principle of the Wiener filter \cite{Doogesh}.
		 \par
		The generation of the $B$-mode by the Thomson scattering in the presence of the tensor perturbation of metric \cite{zal,zal2,zal3,zal4,hu} is the most important method to estimate $r$-parameter. In contrast with the E-mode polarization, the B-mode polarization cannot be generated by the Thompson scattering in the case of the scalar perturbation of metric  \cite{zal,zal2,zal3,zal4, hu,7,8}. The ratio of tensor-to-scalar modes is estimated by comparing the B-mode power spectrum with the E-mode ($r\sim C_{B\,l}/C_{E\,l}$) at least for small $l$s (large-scale).
		  There are several sources such as lensing anomaly \cite{Dome}, vector perturbations \cite{Inomata}, and chiral photons \cite{Inomata1} that mimic signals on the polarization of the CMB. The $B$-mode not only helps to estimate $r$-parameter but also can be used to constrain the bound on the strength of primordial magnetic field \cite{Zucca}, the neutrino masses  \cite{Abazajian}, modification of the gravity \cite{Amendola, Raveri}, cosmic (super-) string \cite{Avgou, Moss, Lizarr}, and other fundamental physics \cite{Abazajian1}. In recent years, many mechanisms have been reported to generate magnetic-like polarization  \cite{Kosowsky:1996yc,Subramanian:2015lua,xue,Scoccola:2004ke,Campanelli:2004pm,Kosowsky:2004zh,Giovannini:2004pf,Bonvin:2014xia,Giovannini:2014bba, Khodagholizadeh:2014nfa, Mohammadi:2015taa,Batebi:2016ocb}. It should be mentioned that small field models of inflation also, can generate a significant primordial gravitational wave signal that can predict the value of $r$-parameter as high as 0.01 \cite{Wolfson1,Wolfson2}.
		\par
		One of the primary sources of curl pattern polarization is  Faraday Rotation, which can provide a distinctive signature of primordial magnetic fields \cite{Pogosian, Sechadri, Kahn}. Magnetic fields generate large vector modes that can be a source for  $B$-mode polarization dominantly, but with the usual thermal CMB power spectrum  \cite{Sechadri,Lewis}. Anisotropic cosmic birefringence can also lead to the conversion of $E$-mode to $B$-mode polarization \cite{Guo}.
		The lensing of the CMB along the line of sight can be another source for $B$-modes polarization, which can be distinguished from the primordial $B$-mode one \cite{lensing}. The vector-mode perturbation due to strings can naturally induce $B$-mode polarization with a spectrum distinct from that expected from inflation itself \cite{Pogosian3}. Also, any instrumental polarization rotation that can convert $E$-mode into $B$-mode and vice versa should be considered \cite{POLO}.\par
		Some of our recent works also discussed the generation of $B$-mode polarization in the presence of scalar perturbations via Cosmic Neutrino Background and CMB interactions \cite{Khodagholizadeh:2014nfa,Khodagholizadeh2,Rohollah}, nonlinear photons interactions \cite{sadegh}, photon interaction by considering extensions to QED such as Lorentz-invariant violating operators \cite{khodam}, non-commutative geometry \cite{Batebi:2016ocb}, interaction of dipolar dark matter with CMB photons \cite{DDM}, and photon-fermion forward scattering  \cite{Zare}. Moreover the intrinsic B-mode polarization is calculated using the Boltzmann code SONG \cite{SONG} that is induced in the CMB by the evolution of primordial density perturbations at the second-order \cite{Wands}.\par
		In our previous work \cite{Vahedi}, we have shown that Compton scattering of photons from polarized electron, which is called Polarized Compton Scattering, can generate circular polarization in contrast to the ordinary Compton scattering  \cite{cosowsky1994}. Nevertheless, we did not investigate the generation of $B$-mode polarization due to polarized Compton scattering, which is the main objective of the present work. In this paper, we discuss the effect of the mentioned mechanism on the amplitude of the primordial gravitational waves ($r$-parameter) and analysis the power spectrum of $E$- and $B$-modes polarization.
		\section{Polarized Cosmic Electrons}
		In the case of Compton scattering of unpolarized in-going electrons (shown by $U_{r}$ spinor state) by photons, one can make an average on initial helicity states of electrons $r$ and an assumption on final states,  which allows using the ordinary completeness relation $\sum_{r}U_{r}(\mathbf{q})\bar{U}_{r}(\mathbf{q})=\frac{q\!\!\!/+m}{m}$. However, in the case of the polarized Compton scattering, we will consider small polarization for in-going electrons. As a result, the Dirac spinors product will be modified to  \cite{kleiss,itzykson}\footnote{It should be mentioned that Eq.(\ref{completeness}) is not completeness relation of Dirac spinors. We do not have any summation over polarization indices of in-going electrons.}
		\begin{eqnarray}\label{completeness}
		U_{r}({q})\bar{U}_{r}({q})=\Big[\frac{q\!\!\!/+m}{2m}\frac{1+\gamma_5{S}\!\!\!/_r(\mathbf{q})}{2}\Big]
		\end{eqnarray}
		where $S_r$ is helicity operator with $r=L,R$ is defined as
		\begin{eqnarray}\label{HO}
		S_R(\mathbf{q})=(\frac{{\mid\mathbf{q}\mid}}{m},\frac{E}{m}\frac{\mathbf{q}}{\mid{\mathbf{q}}\mid}),~~~~~~~~S_L(\mathbf{q})=-S_R(\mathbf{q}).
		\end{eqnarray}
		Let us assume a small fraction $\delta_L$ ($\delta_R$) of left (right)-handed polarization for in-going electrons while we do not apply any constraints on the out-going electrons. \par
		Producing of polarized electrons has been reported in a vast area in physics (See for example \cite{Kessler,Long,Mc,Bell,Kirk}). Here, for example, we address two critical circumstances that inevitably confront us with polarized electrons and thus the asymmetry between left-handed and right-handed electrons would happen. The presence of an external magnetic field makes electrons  occupy Landau levels and beta-processes in Big Bang Nucleosynthesis (BBN), which make a discrepancy in the interaction of left- and right- handed electrons with left-handed neutrino.
		\subsection{At the presence of  primordial magnetic field }
		It is believed that the early universe was filled with high conductivity charged plasma. According to this theory, the universe might have possessed a stochastic magnetic field that was in a dynamical co-evolution with expanding matter  \cite{Durrer:2013pga}.  From the study of quadrupole anisotropy in CMB, one can justify that a very large-scale field such as a magnetic field would select out a particular direction  \cite{Thorne67}. Nevertheless, the origin of the primordial magnetic field is a challenging question that has attracted much interest in the physics community (for more information, see \cite{Subramanian:2015lua} and references therein). Here, we review the effect of the possible external cosmic magnetic field on the generation of polarized cosmic electrons. \par
		Energy spectrum of the left-handed and right-handed fermions field through the Dirac equation at the presence of a constant magnetic field along the $z$-direction, would be
		\begin{equation}\label{eq:db}
		E_n=\pm \sqrt{m^2+p_z^2+2n\, e\, B} \,,\qquad n=0,\pm 1,\pm 2,\dots ,
		\end{equation}
		where $n$ counts Landau levels. It has to be noted that after the last scattering, the cosmic electrons are non-relativistic particles and then for such non-relativistic electrons, we have,
		\begin{equation}\label{eq:ndb}
		E_n \approx \frac{p_z^2}{2 m}+\frac{n\,e B}{m}.
		\end{equation}
		The exciting phenomenon will happen at the lowest Landau level $n=0$. At this level, at least, there is no symmetry in the occupation between left- and right-handed charged fermions (see  \cite{Bhattacharya:2007vz} for the detailed discussion).
		 Consider the cosmic electrons as a fermionic gas with $N$ particles with the energy as Eq.\eqref{eq:ndb}.  It is clear that  $E_F\ge \frac{n e B}{m}$ where $E_F$ is Fermi energy. The equality will  happen with maximum Landau level $n_{max}$ as follows
		\begin{equation}
		n_{max}=\frac{E_F}{\frac{ e B}{m}}
		\end{equation}
		So, one can consider an asymmetry to left- and right- handed electrons as
		\begin{equation}\label{deltal}
		\delta_L\sim \frac{1}{n_{max}}=\frac{e B}{m E_F}\,.
		\end{equation}
		The dependence of $\delta_L$ (due to magnetic field) to red-shift is another issue which we need to discuss. To investigate the mention issue, we start with the evolution of primordial magnetic field and the density of cosmic electrons  during universe expansion. 
		Following~\cite{Subramanian:2015lua}, the value of the primordial magnetic field and cosmic baryon density, as well as electron density, in terms of red-shift are given as
		\begin{equation}
		B_0=B(t_0)(1+z)^2,\,\,\,\,\,\,\,\,\, n_e\sim n_b=n_b(t_0)(1+z)^3
		\end{equation}
		where $z$ is a red-shift parameter.\\
		The Fermi energy for cosmic electrons in the non-relativistic three-dimensional system can be written as
		\begin{equation}
		E_F=\frac{(3\pi^2 n_b )^{2/3}}{2 m},
		\end{equation}
		Therefore, from Eq. \eqref{deltal}, $\delta_L$ coming from the primordial magnetic field is almost independent of the red-shift and it  would take  the same value in all universe scales. Finally, by considering  $n_e(t_0)=n_b(t_0)\simeq 10^{-7} (\frac{1}{cm^3})$ for the present density of cosmic electron, we have
		\begin{equation}
		\delta_L\approx 10^{-4} B_{18},
		\end{equation}
		where  $B_{18}=B/10^{-18}G$.
Note the primordial magnetic field, in large scale, is a stochastic field. Despite this fact, our above arguments remain credible because the asymmetry in occupation between left- and right-handed charged fermions in Lowest Landau Level is independence of the direction of magnetic field, see for example \cite{Bhattacharya:2007vz}.
		\subsection{Beta process in BBN}
		One of the most important parameters to study during BBN is the neutron-proton number ratio. The neutron-proton ratio was estimated by Standard Model physics before the nucleosynthesis epoch; almost the first second after the Big Bang. Before the nucleosynthesis era, the neutron-proton ratio ($\frac{n}{p}$) was close to $1$. At the freeze-out period, this ratio would be $\frac{1}{6}$ and after freeze-out gets smaller. \par
		It is well known that neutrinos interact with electrons and nucleons via charged and neutral current while the charged current, $\beta$ process, is dominated. In addition, due to the parity-violating coupling of neutrinos to matter, neutrinos interacting only with left-handed quarks and electrons by exchanging  charged gauge bosons $W^\pm$. However left-handed neutrino can be coupled to left- and right-handed quarks ($u,d$) by exchanging neutral gauge boson $Z^\circ$,
		\begin{eqnarray}
		n+\nu_{eL}    &\rightarrow & p+e^{-}_{L} \\
		n+e^{+}    &\rightarrow & p + \tilde{\nu}_e
		\end{eqnarray}
		This fact can be a source to generate the asymmetry between left- and right-handed polarization of cosmic electrons. Although the neutrons react through the above reactions to produce protons and polarized electrons, these polarized electrons can make secondary interactions (during the time between freeze out to last scattering epoch) to lose their polarization.
		It has to be noted that, in this paper, we do not study these effects exactly (may happen in future) and we just mention it as our motivation.
		\section{Power Spectrum of Scalar modes in presence of Polarized
			Compton Scattering}
		To get the time evolution of CMB polarization, using the quantum Boltzmann equation is helpful, especially when we need to consider different collision terms. Such an approach was studied in \cite{cosowsky1994}. The Boltzmann equation for CMB polarization via ordinary and polarized Compton scattering is derived in \cite{Vahedi} (see also Appendix). In the following, we just consider the equation for linear polarization, which is given as
		\begin{eqnarray} \label{Boltzmann}
		\frac{d}{d\eta}\Delta_P^{\pm(S)} +iK\mu \Delta_P^{\pm(S)}
		&=&\dot{\tau}_{e\gamma} [-\Delta_P^{\pm(S)}+\dfrac{1}{2}(1-P_{2}(\mu))\Pi]\nonumber\\
		&\pm& i \dot{\tau}_{_{\rm PC}} \dfrac{2}{3}\Delta_{\rm{I2}}^{(S)}(1-\mu^{2})\pm\dfrac{1+i}{3}(1-\mu^{2})\dot{\tau}_{_{\rm PC}}\Pi^\pm
		\end{eqnarray}
		where $\Delta^{\pm(S)}_{P}(\mathbf{K},\mathbf{k},\tau)=Q^{(S)}\pm iU^{(S)}$, $Q$ and $U$ are Stokes parameters to describe linear polarization,  $(S)$ indicates the primordial scalar perturbations which is expanded in the Fourier modes characterized by wave number $\mathbf{K}$, $\dot{\tau}_{e\gamma}\equiv \frac{d\tau_{e\gamma}}{d\eta}$ is Compton scattering optical depth, $a(\eta)$ is normalized scale factor,  $\mu = \hat{n}\cdot\hat{\mathbf{K}} = \cos \theta$, where $\theta$ is the angle between the CMB photon direction $\hat{n}=\mathbf{k}/|\mathbf{k}|$ and the wave vectors
		$\mathbf{K}$, and $P_2(\mu)$ is the Legendre polynomial of rank $2$. In equation (\ref{Boltzmann}), the source terms $\Pi\equiv \Delta_{\rm{I2}}^{(S)}+\Delta_{\rm{Q2}}^{(S)}+\Delta_{\rm{Q0}}^{(S)}$ comes from usual Compton scattering while the source term from polarized Compton is
		\begin{eqnarray}
		\Pi^{+}
		&=&(2+i)\Delta_{\rm{P2}}^{+(S)}+i\Delta_{\rm{P2}}^{-(S)},\,\,\,\,\,\label{Pi+}\\
		\Pi^{-}&=&(2i+1)\Delta_{\rm{P2}}^{-(S)}+\Delta_{\rm{P2}}^{+(S)},\,\,\,\,\,
		\end{eqnarray}
		where
		\begin{equation}\label{optical-pc1}
		\dot{\tau}_{_{\rm PC}}=\frac{3}{2} \frac{m\,v_e(\mathbf{x})}{k^0}  \sigma_T n_{eL}(\mathbf{x})=\frac{3}{2} \frac{m\,v_e(\mathbf{x})}{k^0}  \sigma_T \delta_{\rm{L}} n_{e}(\mathbf{x}),
		\end{equation}
		where $v_e(\mathbf{x})$ is electron bulk velocity. Note that the sources in the above equations involve the multipole moments of intensity $I$ and polarization $P$, defined as $\Delta^{(S)}(K,\mu)=\sum_l(2l+1)(-i)^{l}\Delta^{(S)}_l(K)P_l(\mu)$,
		where $P_l(\mu)$ is the Legendre polynomial of order $l$.\par
		The value of $\Delta _{\rm{P}}^{ \pm(S)}(\hat{n})$  at
		the present time $\eta_0$ and the direction $\hat{n}$ can be obtained in the following general form  by integrating of the Boltzmann equation (\ref{Boltzmann}) along the line of sight  \cite{zal} and summing over all the Fourier modes $\mathbf{K}$ as follows
		\begin{eqnarray}
		\Delta _{\rm{P}}^{\pm(S)}(\hat{\bf{n}})
		&=&\int d^3 \bf{K} \xi(\bf{K}) e^{ 2i \varphi_{\mathbf{K},n}}\Delta _{\rm{P}}^{ \pm(S)}
		(\mathbf{K},\mathbf{k},\eta_0),\,\,\,\,\,\label{Boltzmann03}
		\end{eqnarray}
		where $  \varphi_{\mathbf{K},n}$  is the angle needed to rotate the $ \mathbf{K} $ and $ \mathbf{n} $ dependent basis to a fixed frame in the sky  and $\xi(\mathbf{K})$ is a random value  that is used  to
		characterize the initial amplitude of each primordial scalar perturbations mode. Here,the values of
		$\Delta _{\rm{P}}^{\pm (S)}(\mathbf{K},\mathbf{k},\eta_0)$ are given as
		\begin{eqnarray} \label{deltap}
		\Delta _{\rm{P}}^{\pm (S)}
		(\mathbf{K},\mu,\eta_0)&=&\int_0^{\eta_0} d\eta\,\dot\tau_{e\gamma}\,
		e^{ix \mu -\tau_{e\gamma}}\,\,\Big[\frac {3}{4} (1-\mu^2)\Pi(K,\eta) \pm i \dfrac{2 \dot{\tau}_{_{\rm PC}}}{3\dot\tau_{e\gamma}}\Delta_{\rm{I2}}^{(S)}(1-\mu^{2}) \nonumber\\&&~~~~~~~~~~~~~~~ \ \pm \dfrac{1+i}{3}(1-\mu^{2})\dfrac{\dot{\tau}_{_{\rm PC}}}{\dot\tau_{e\gamma}} \Pi^{\pm}\Big]
		,\label{EBS}
		\end{eqnarray}
		where $x=K(\eta_0 - \eta)$. Differential optical depth  $\dot\tau_{e\gamma}(\eta)=a\,n_e\,\sigma_T$  and total optical depth $\tau_{e\gamma}(\eta)$ due to the Thomson scattering at time  $\eta$ are defined as
		\begin{equation}\label{optical}
		\tau_{e\gamma}(\eta)=\int_\eta^{\eta_0}\dot{\tau}_{e\gamma}(\eta) d\eta.
		\end{equation}
		As it is well known, the linear polarization of CMB can be described in terms of the divergence-free part (B-mode $ \Delta_{\rm{B}}^{(S)} $) and  the curl-free part (E-mode $ \Delta_{\rm{E}}^{(S)} $) instead of $Q$ and $U$ parameters as below
		\begin{eqnarray}
		\Delta_{\rm{E}}^{(S)}(\hat{n})=-\dfrac{1}{2}[\bar{\eth}^{2}\Delta_{\rm{P}}^{+(S)}+\eth^{2}\Delta_{\rm{P}}^{ -(S)}]\nonumber\\
		\Delta_{\rm{B}}^{(S)}(\hat{n})=\dfrac{i}{2}[\bar{\eth}^{2}\Delta_{\rm{P}}^{+(S)}-\eth^{2}\Delta_{\rm{P}}^{ -(S)}]
		\end{eqnarray}
		where $\eth$ and $\bar{\eth}$ are spin raising and lowering operators, respectively, \cite{zal}. In $ \vec{K}\| z $ coordinate frame and considering azimuthal symmetry give $\bar{\eth}^{2} \equiv \eth^{2}$.
		Finally, the power spectrum of linear polarization in CMB, $C_{\rm{X}\,l}^{(S)}$ because of a general interaction in the presence of scalar perturbation is given by the equation (\ref{clx})
		\begin{eqnarray}\label{clx}
		C_{\rm{X}\,l}^{(S)}=\dfrac{1}{2l+1}\sum_{m}\langle a_{X, lm}^{\ast}a_{X,lm}\rangle
		\end{eqnarray}
		where $ X = \lbrace \rm{E}\,, \rm{B} \rbrace $ and
		\begin{eqnarray}
		a_{\rm{E}\,lm}=\int d\Omega ~Y_{lm}^{\ast} ~\Delta_{\rm{E}}\,,~~~~~~~~~~~~    a_{\rm{B}\,lm}=\int d\Omega ~Y_{lm}^{\ast} ~\Delta_{\rm{B}}.
		\end{eqnarray}
		In the following, we report the effect of polarized Compton scattering on E- and B-modes power spectrum.
		
		\begin{figure}
			\begin{center}
				\includegraphics[scale=0.60]{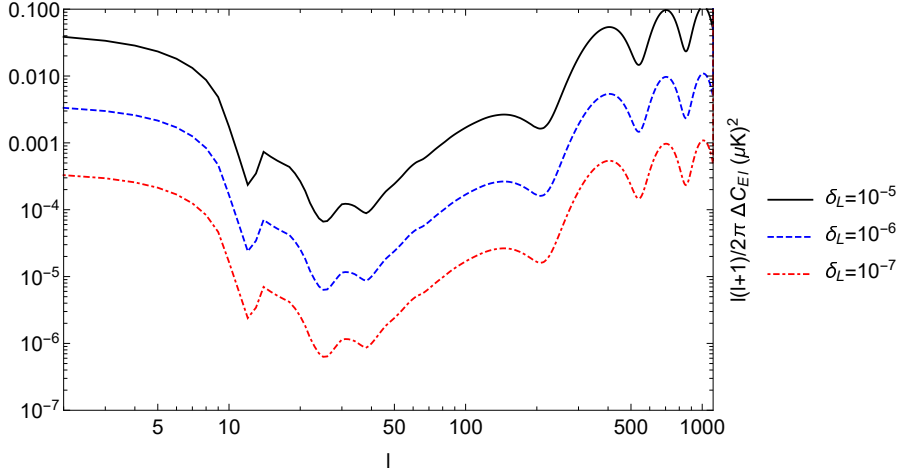}
			\end{center}
			\caption{ The deviation plot of E-mode power spectrum  from standard one via polarized Compton scattering in the presence of scalar perturbation for different $ \delta_{L}$.}
			\label{DeltaClE}
		\end{figure}
		\subsection{E-Mode in presence of Polarized Compton Scattering}
		By considering the polarized Compton scattering,   the modified Boltzmann equation ( equation (\ref{Boltzmann})) with  acting the spin raising operator twice on the integral solution of $ \Delta _{P}^{\pm (S)}
		(\mathbf{K},\mu,\eta_0)$ (equation (\ref{deltap})) leads to the following expressions for electric-like polarization in the presence of the scalar perturbations
		\begin{eqnarray}
		\Delta_{\rm{E}}^{(S)}(\eta_{0},k,\mu)&=&-\int_{0}^{\eta_{0}}d\eta  g(\eta) [\dfrac{3}{4} \Pi(K,\eta)+\dfrac{2}{3}\Delta_{\rm{P2}}^{(S)}(K,\eta) \dfrac{\dot{\tau}_{_{\rm PC}}}{\dot{\tau}_{_{ \rm e\gamma}}}] \partial_{\mu}^{2}[(1-\mu^{2})^{2} e^{ix\mu}]\nonumber\\ &=& \int_{0}^{\eta_{0}}d\eta g(\eta) [\dfrac{3}{4} \Pi(K,\eta)+\dfrac{2}{3}\Delta_{\rm{P2}}^{(S)}(K,\eta) \dfrac{\dot{\tau}_{_{\rm PC}}}{\dot{\tau}_{_{ \rm e\gamma}}}]  (1+\partial_{x}^{2})^{2} (x^{2} e^{ix\mu})
		\end{eqnarray}
		Therefore, the E-mode power spectrum $C_{El}^{(S)}$ due to  polarized Compton scattering in addition the ordinary Compton scattering in the presence of scalar perturbation background would be
		\begin{eqnarray}
		C_{\rm{E}\,l}^{(S)}&=&\dfrac{1}{2l+1}\sum_{m}\langle a_{\rm{E},lm}^{\ast}a_{\rm{E},lm}\rangle \nonumber\\
		&=& \dfrac{1}{2l+1}\dfrac{(l-2)!}{(l+2)!} \int d^{3} \vec{K} P_{\varphi}^{(S)}(\vec{K},\tau)\sum_{m} \vert \int  d\Omega Y_{lm}^{\ast} \int_{0}^{\eta_{0}}d\eta  g(\eta)  [\dfrac{3}{4} \Pi(K,\eta)+\dfrac{2}{3}\Delta_{\rm{P2}}^{(S)}(K,\eta) \dfrac{\dot{\tau}_{_{\rm PC}}}{\dot{\tau}_{_{ \rm e\gamma}}}]  \nonumber\\ &&~~~~~~~~~~~~~~~~~~~~~~~~~~~~~~~~~~~~~~~~~~~~~~~~~~~~~~~~~~~~~~~~~~\times [ (1+\partial_{x}^{2})^{2} (x^{2} e^{ix\mu} ) ] \vert^{2}
		\end{eqnarray}
		\begin{figure}
			\begin{center}
				\includegraphics[scale=0.7]{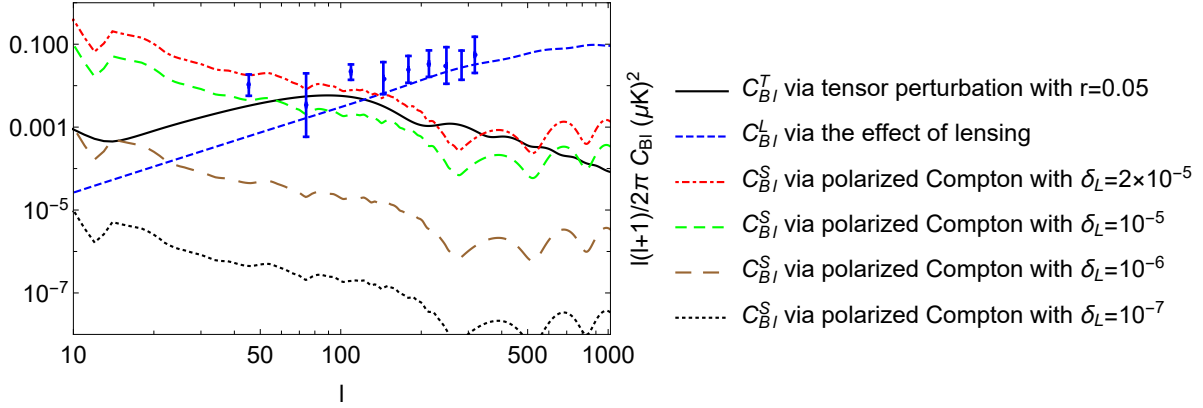}
			\end{center}
			\caption{B-mode power spectrum plotted in terms of $(\mu K)^2$ for different cases; here, $C^T_{B l}$ indicates the contribution of Compton scattering in the presence of tensor perturbations with tensor-to-scalar ratio $r=0.05$ while $C^L_{B l}$ indicates the Lesing contribution. Also, $C^S_{B l}$ is the distribution of polarized Compton in the presence of scalar perturbations. To compare the results, the experiment BICEP2/Keck/Plank 2018 results for the B-mode power spectrum (dots with their error bars) were added.}
			\label{ClB}
		\end{figure}
		Considering the $ \vec{K} || z $ condition, we would have  $ \int d\Omega Y_{lm}^{\star}(\hat{n}) \,e^{ix\mu}=\sqrt{4\pi(2l+1)}\, i^{l}  \,j_{l}(x)\,\delta_{m0}$ and the differential equation satisfied by the spherical Bessel function, $ j_{l}^{''}(x)+2\dfrac{j_{l}^{'}(x)}{x} +[1-\dfrac{l(l+1)}{x^{2}}]\, j_{l}(x)=0$, hence, the E-mode power spectrum could be rewritten as
		\begin{eqnarray}
		C_{\rm{E}\,l}^{(S)}&=& (4\pi)^{2}\dfrac{(l+2)!}{(l-2)!}\int K^{2} dK P_{\varphi}(K) \Big(\int_{0}^{\eta_{0}} d\eta  g(\eta) [\dfrac{3}{4} \Pi(K,\eta)+\dfrac{2}{3}\Delta_{\rm{P2}}^{(S)}(K,\eta) \dfrac{\dot{\tau}_{_{\rm PC}}}{\dot{\tau}_{_{ \rm e\gamma}}}] \dfrac{j_{l}(x)}{x^{2}}\Big)^{2}\nonumber\\  &\simeq&
		(4\pi)^{2}\dfrac{(l+2)!}{(l-2)!}\int K^{2} dK P_{\varphi}(K) \Big\{\Big(\int_{0}^{\eta_{0}} d\eta  g(\eta) \dfrac{3}{4} \Pi(K,\eta) \dfrac{j_{l}(x)}{x^{2}}\Big)^{2} \nonumber\\  ~~~~~~~~~~~~~~~&+& \int_{0}^{\eta_{0}} d\eta  g(\eta) \Pi(K,\eta)\Delta_{\rm{P2}}^{(S)}(K,\eta) \dfrac{\dot{\tau}_{_{\rm PC}}}{\dot{\tau}_{_{ \rm e\gamma}}}\Big(\dfrac{j_{l}(x)}{x^{2}}\Big)^2\Big\}\label{deltaClE1}
		\end{eqnarray}
		The first term in the second line of the above equation presents the value of the E-mode power spectrum from the standard scenario of cosmology $\bar{C}_{\rm{E}\,l}^{(S)}$ and the second term comes from the Polarized Compton scattering. Note we neglect the term  inclouding  $(\dfrac{\dot{\tau}_{_{\rm PC}}}{\dot{\tau}_{_{ \rm e\gamma}}})^2$.  Therefore, deviation E-mode power spectrum from their standard value, $\Delta C_{\rm{E}\,l}^{(S)} $, can be written as
		\begin{eqnarray}
		\Delta C_{\rm{E}\,l}^{(S)}=C_{\rm{E}\,l}^{(S)}-\bar{C}_{\rm{E}\,l}^{(S)}. \end{eqnarray}
		Therefore from Eq. (\ref{deltaClE1}) we can show, 
		\begin{eqnarray}
		\Delta C_{\rm{E}\,l}^{(S)}\propto \bar{\eta}\,\bar{C}_{\rm{E}\,l}^{(S)}, \label{propto1} \end{eqnarray}
		where $\bar\eta$ is the time average value of $ \frac{\dot{\tau}_{\rm PC}}{\dot{\tau}_{e\gamma}}$. This means that the behavior of $\Delta C_{\rm{E}\,l}^{(S)}$ is more or less similar to $\bar{C}_{\rm{E}\,l}^{(S)}$ which is oscillated by $l$ and peaked around $1000<l<1500$. For this reason, we have just plotted the deviation of E-mode power spectrum from standard one via polarized Compton scattering for $l<1500$, however as mention we expect an oscillating behavior for this quantity in the case $l>1500$. 
		In Fig. \ref{DeltaClE} for different $\delta_L$ values, $\Delta C_{\rm{E}\,l}^{(S)}$ is plotted in terms of $l$. As this plot shows, $\Delta C_{E l}$ for $\delta_{L}=10^{-5}$ is in the order of $10^{-3}\, (\mu K)^2$, at least for small $l$s, which is in the range of the current precision experiments.
		\begin{figure}
			\begin{center}
				\includegraphics[scale=0.7]{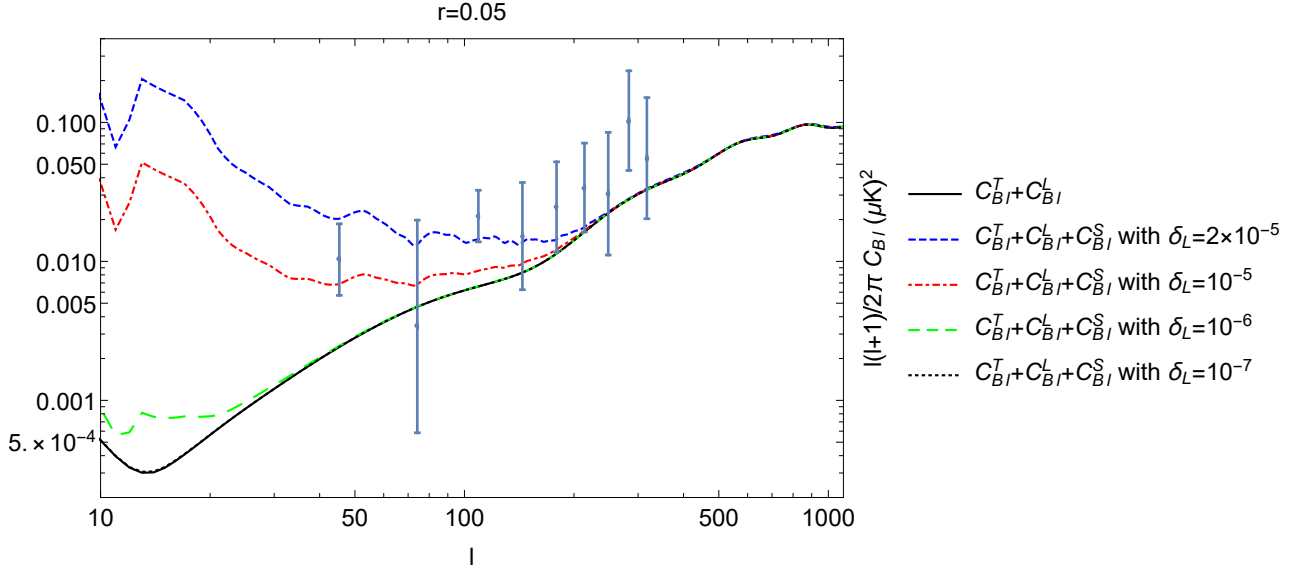}\end{center}
			\caption{ By considering the lensing effects, tensor perturbations and polarized Compton scattering effects in the presence of scalar perturbations, the total value of the B-mode power spectrum for the different value of $\delta_L$ are plotted. Also, the experiment BICEP2/Keck/Plank 2018 results for B-mode power spectrum (dots with their error bars) were added. The Black line is the B-mode power spectrum is tensor perturbation along with the lensing effect (standard value).  Also, by adding the B- mode power spectrum obtained in the presence of scalar perturbation for different $\delta_{L}$ is plotted: the blue line is for $ \delta_{L}=2\times 10^{-5} $, the red line is for $ \delta_{L}=10^{-5}$, the green line is for $ \delta_{L}= 10^{-6}$ and the black dotted line is for $ \delta_{L}= 10^{-7}$.}
			\label{ClB-Total}
		\end{figure}
		\subsection{B-Mode power spectrum}
		In the standard scenario of cosmology for CMB polarization, by considering azimuthal symmetry, we have $\bar{\eth}^{2} \Delta_{\rm{P}}^{+(S)}=\eth^{2} \Delta_{\rm{P}}^{-(S)}$, therefore  $\Delta_{\rm{B}}^{(S)}(\eta_{0}, k, \mu) $ would be zero. In the presence of scalar perturbation, the B-mode can not be generated via ordinary Compton scattering  $\bar{C}_{\rm{B}\,l}^{(S)}=0 $. However, considering the contribution of polarized Compton scattering, our result leads to the following expression
		\begin{eqnarray}
		\tilde{\Delta}_{\rm{B}}^{(S)}(\eta_{0},k,\mu)&=&\dfrac{2}{3} \int_{0}^{\eta_{0}}d\eta g(\eta)\, [\Delta_{\rm{I2}}^{(S)}(K,\eta)+(4i-1)\Delta_{\rm{P2}}^{(S)}(K,\eta)]\, \dfrac{\dot{\tau}_{_{\rm PC}}}{\dot{\tau}_{_{ \rm e\gamma}}} \partial_{\mu}^{2}\,((1-\mu^{2})^{2} e^{ix\mu})\nonumber\\ &=& -\dfrac{2}{3}\int_{0}^{\eta_{0}}d\eta g(\eta) \dfrac{\dot{\tau}_{_{\rm PC}}}{\dot{\tau}_{_{ \rm e\gamma}}}\, [\Delta_{\rm{I2}}^{(S)}(K,\eta)+(4i-1)\Delta_{\rm{P2}}^{(S)}(K,\eta)] \, (1+\partial_{x}^{2})^{2} (x^{2} e^{ix\mu})\nonumber\\
		\end{eqnarray}
		Therefore, the B-mode power spectrum, $ C_{\rm{B}\,l}^{(S)} $, would be
		\begin{eqnarray}
		C_{\rm{B}\,l}^{(S)}&=&\bar{C}_{\rm{B}\,l}^{(S)}+\Delta C_{\rm{B}\,l}^{(S)}=\dfrac{1}{2l+1}\sum_{m}\langle a_{\rm{B}\, lm}^{\ast}a_{\rm{B}\,lm}\rangle \nonumber\\
		&=& \dfrac{1}{2l+1}\dfrac{(l-2)!}{(l+2)!} \int d^{3} \vec{K} P_{\varphi}^{(S)}(\vec{K},\tau)\sum_{m} \Big\vert\frac{2}{3} \int  d\Omega Y_{lm}^{\ast}  \nonumber\\ &&\int_{0}^{\eta_{0}}d\eta  g(\eta) [\Delta _{\rm{I2}}^{(S)}(K,\eta)+(4i-1)\Delta_{\rm{P2}}^{(S)}(K,\eta)]\, \dfrac{\dot{\tau}_{_{\rm PC}}}{\dot{\tau}_{_{ \rm e\gamma}}}\, [ (1+\partial_{x}^{2})^{2} (x^{2} e^{ix\mu} ) ] \Big\vert^{2}.
		\end{eqnarray}
		Finally, the B-mode power spectrum because of polarized Compton scattering in the presence of scalar perturbation can be written as
		\begin{eqnarray}
		C_{\rm{B}\,l}^{(S)}&=&\Delta C_{\rm{B}\,l}^{(S)}= (4\pi)^{2}\dfrac{(l+2)!}{(l-2)!}\int K^{2} dK P_{\varphi}(K) \nonumber\\
		&&\Big(\dfrac{2}{3}\int_{0}^{\eta_{0}} d\eta  g(\eta) \dfrac{\dot{\tau}_{_{\rm PC}}}{\dot{\tau}_{_{\rm e\gamma}}} [\Delta _{\rm{I2}}^{(S)}(K,\eta)+(4i-1)\Delta_{\rm{P2}}^{(S)}(K,\eta)] \dfrac{j_{l}(x)}{x^{2}}\Big)^{2}.\label{ClBS1}
		\end{eqnarray}
		The effect of polarized Compton scattering on a tensor-to-scalar ratio ($r$-parameter) and the B-mode  power spectrum can not be ignored. As mentioned, several times, in the standard scenario of cosmology, we have $\bar{C}_{\rm{B}\,l}^{(S)}=0$. From this equation, we have $C_{\rm{B}\,l}^{(Ob)}=C_{\rm{B}\,l}^{(T)}+C_{\rm{B}\,l}^L$, where $C_{\rm{B}\,l}^{(Ob)}$ indicates the observed B-mode power spectrum and $C_{\rm{B}\,l}^{L}$ is B-mode power spectrum generated by the lensing effects  while $C_{\rm{B}\,l}^{(T)}$ is B-mode power spectrum due to ordinary Compton scattering in the presence of gravitational wave. As a result, we could write the standard value of the tensor-to-scalar ratio $r$ as follows
		\begin{eqnarray}
		r=P_{T}/P_{S}\propto C_{\rm{B}\,l}^{(T)}/C_{\rm{E}\,l}^{(S)}\simeq C_{\rm{B}\,l}^{(Ob)}/C_{\rm{E}\,l}^{(S)}, \label{r0}
		\end{eqnarray}
		here we neglect $C_{\rm{B}\,l}^{L}$ which has small contribution for small $l$.\footnote{ Note Eq.(\ref{r0}) is not precise equation to calculate $r-$parameter, for more detail see \cite{rr1,rr2}, we just use this equation to give a sense about the effect of polarized Compton scattering on $r-$parameter value.} But in our case, $C_{\rm{B}\,l}^{(S)}\neq0$, the observed B-mode power spectrum is $C_{\rm{B}\,l}^{(Ob)}=C_{\rm{B}\,l}^{(S)}+C_{\rm{B}\,l}^{(T)}$. So, we have
		\begin{eqnarray}\label{rstar}
		r^*\simeq C_{\rm{B}\,l}^{(T)}/C_{\rm{E}\,l}^{(S)}=r-\dfrac{C_{\rm{B}\,l}^{(S)}}{C_{\rm{E}\,l}^{(S)}},\label{r1}
		\end{eqnarray}
		where we call  $r^*$ as a {\it net} scalar-to-tensor ratio. From equations (\ref{rstar}) and (\ref{ClBS1}), we can yield the below result
		\begin{eqnarray}
		r^*\simeq r-\Big(\,\overline{\dfrac{\dot{\tau}_{_{\rm PC}}}{\dot{\tau}_{_{\rm e\gamma}}}}\,\Big)^2,\label{r2}
		\end{eqnarray}
		where
		\begin{eqnarray}
		\overline{\dfrac{\dot{\tau}_{_{\rm PC}}}{\dot{\tau}_{_{\rm e\gamma}}}}=\dfrac{1}{\eta_0}\int_0^{\eta_0}\dfrac{\dot{\tau}_{_{\rm PC}}}{\dot{\tau}_{_{\rm e\gamma}}}\simeq\,10^{-3}\,\Big(\dfrac{\delta_L}{10^{-7}}\Big).\label{r3}
		\end{eqnarray}
		Finally, we can estimate the {\it net} scalar-to-tensor ratio as follows
		\begin{eqnarray}\label{rprim}
		r^*\simeq r-\,10^{-6}\,\Big(\dfrac{\delta_L}{10^{-7}}\Big)^2.\label{r4}
		\end{eqnarray}
		As can be seen from equation (\ref{rprim}), the contamination from polarized Compton scattering can be comparable to a primordial tensor-to-scalar ratio spatially for $\delta_L>10^{-5}$.
		\section{Conclusion}
		In this paper, first, we shortly investigate the asymmetry in the number density of left- and right-handed cosmic electrons ($\delta_L$ and $\delta_R$, respectively) due to the primordial large-scale magnetic field and beta processes in BBN epoch. Next, by solving the quantum Boltzmann equation, the time evolution of Stokes parameters via ordinary (unpolarized) and polarized Compton scattering is obtained. We have shown that the polarized Compton scattering, in contrast with the ordinary one, can generate a magnetic-like pattern in linear polarization of CMB radiation. We have also shown that the B- mode power spectrum of CMB in the presence of scalar perturbation does not vanish and its value depends on the square value $\delta_L^2$ ($C_{B\,l}^{(S)}\propto \delta_L^2$). We have plotted the power spectrum of the B-mode generated by the polarized Compton scattering and we have compared it with the power spectra produced by weak lensing effects and Compton scattering in the presence of tensor perturbations (Figs. (\ref{ClB}-\ref{ClB-Total})). The results show a significant amplification in $C_{B\,l}$ in large scale $l<500$ for $\delta_L>10^{-6}$, which can be observed in future high-resolution B-mode polarization detection. Also, we showed that $C_{ B\,l}^{(S)}$ generated by polarized Compton scattering can suppress the tensor-to-scalar ratio, $r$ parameter, so that the contamination from polarized Compton scattering may be comparable to a primordial tensor-to-scalar ratio spatially for $\delta_L>10^{-5}$.
		\section{\small Acknowledgment}
		A. Vahedi would like to thank S. Khosravi, F. Kanjouri, and M. Afkani for help during the numerical calculation.

		
		\section*{Appendix: CMB Interactions with Polarized Electrons }
		The effects of the external magnetic field on a large scale \cite{cooray}, chiral magnetic instability in neutron stars and Magnetars \cite{ohnishi}, fermion production during and after axion inflation \cite{adshead},  and new physics interactions on the distribution of cosmic electrons can be considered as possible sources of the polarized cosmic electrons. These effects motivated us to consider the generation of CMB circular polarization via polarized Compton scattering.
		Recently,  in \cite{Vahedi}, by straightforward calculating the interaction Hamiltonian  for photon-polarized electron scattering ($ e+\gamma\longrightarrow e+\gamma $), the Boltzmann equation for  $  \rho_{ij}(\mathbf{x},\mathbf{k}) $  in the first order of the interaction Hamiltonian was presented as
		\begin{eqnarray}\label{cegamma}
		\frac{d}{dt}\rho_{ij}(\mathbf{x},\mathbf{k})&&=\frac{e^4}{2k^0(2k.q)^2}(i)\int d\mathbf{q}d\mathbf{p}\frac{m}{E(\mathbf{q}+\mathbf{k}-\mathbf{p})}(2\pi)\delta\big(E(\mathbf{q}+\mathbf{k}-\mathbf{p})+p-E(\mathbf{q})-k\big)\nonumber\\
		&&\times\bigg(n_{eL}(\mathbf{x}, \bold q)\delta_{s_2s'_1}(\delta_{is_1}\rho_{s'_2j}(\mathbf{k})+\delta_{js'_2}\rho_{is_1}(\mathbf{k}))-2n_{eL}(\mathbf{x}, \bold q')\delta_{is_1}\delta_{js'_2}\rho_{s'_1s_2}(\mathbf p)\bigg)\mid\mathcal M\mid_P^{~~2} \nonumber\\
		\end{eqnarray}
		where $\bold q$, $\bold p$, $\bold k$ and $n_{eL}(\mathbf{x},\bold q')$ are incoming electron momentum, incoming photon momentum, outgoing photon momentum of Compton scattering amplitude, and number density of polarized cosmic electrons respectively. We consider $\hat{q}=\vec q/\mid q\mid$ and   $ \mid\mathcal M\mid_P^{~~2}  $ as the contribution of Compton scattering of photons by polarized electrons as
		\begin{eqnarray}\label{mp2}
		\mid\mathcal M\mid_P^{~~2}\approx \frac{e^4}{4(q.k)^2}\Bigg\{&&q.\epsilon_{s'_2}(k)\Big(k.\epsilon_{s'_1}(p)\hat{q}.\epsilon_{s_1}(k)\times\epsilon_{s_2}(p)+p.\epsilon_{s_1}(k)\hat{q}.\epsilon_{s'_1}(p)\times\epsilon_{s_2}(p)\Big)\nonumber\\
		&&+q.\epsilon_{s_2}(p)\Big(p.\epsilon_{s_1}(k)\hat{q}.\epsilon_{s'_2}(k)\times\epsilon_{s'_1}(p)+\hat{q}.\epsilon_{s_1}(k)\epsilon_{s'_2}(k).p\times\epsilon_{s'_1}(p)\Big)\nonumber\\
		&&+\hat{q}.\epsilon_{s'_1}(p)\Big(q.\epsilon_{s_2}(p)k.\epsilon_{s_1}(k)\times\epsilon_{s'_2}(k)-q.\epsilon_{s'_2}(k)\epsilon_{s_2}(p).k\times\epsilon_{s_1}(k)\Big)\nonumber\\
		&&-q.\epsilon_{s'_2}(k)\hat{q}.\epsilon_{s_1}(k)p.\epsilon_{s'_1}(p)\times\epsilon_{s_2}(p)\nonumber\\
		&&+\epsilon_{s_1}(k).\epsilon_{s'_1}(p)\Big(q.\epsilon_{s_2}(p)\hat{q}.k\times\epsilon_{s'_2}(k)-q.\epsilon_{s'_2}(k)\hat{q}.k\times\epsilon_{s_2}(p)\nonumber\\
		&&+q.\epsilon_{s_2}(p)\hat{q}.p\times\epsilon_{s'_2}(k)-q.\epsilon_{s'_2}(k)\hat{q}.p\times\epsilon_{s_2}(p)\Big) \nonumber\\
		&&+\epsilon_{s_1}(k).\epsilon_{s_2}(p)q.\epsilon_{s'_2}(k)\hat{q}.p\times\epsilon_{s'_1}(p)+\epsilon_{s'_1}(p).\epsilon_{s'_2}(k)q.\epsilon_{s_2}(p)\hat{q}.k\times\epsilon_{s_1}(k)\nonumber\\
		&&-\delta_{s_2s'_1}q.\epsilon_{s'_2}(k)\hat{q}.k\times\epsilon_{s_1}(k)-\delta_{s_1s'_2}q.\epsilon_{s_2}(p)\hat{q}.p\times\epsilon_{s'_1}(p)\Bigg\}
		\end{eqnarray}
		where $\epsilon_{s_1}$  is the polarization vector component of incoming and outgoing photons.  By running  all indices and defining equation (\ref{mp2}) as vector-like object $\mathcal{M}_P(s_1,s_2,s'_1,s'_2)$ and doing integration over $\bold q$ and spatial integration over $\bold p$, the main Stokes parameters take the following form
		\begin{eqnarray}\label{idotk}
		\dot{I}(\mathbf{k})&&=\dfrac{1}{2}(\dot{\rho}_{11}+\dot{\rho}_{22}) \nonumber \\ &&= i\dot{\tau}_{_{\rm PC}}\int \frac{d\Omega}{4\pi}\bigg[f_{II}{(\hat k,\hat p)}I(\bold k)+f_{IQ}{(\hat k,\hat p)}Q(\bold k)+f_{IU}{(\hat k,\hat p)}U(\bold k)+ f_{IV}{(\hat k,\hat p)}V(\bold k)\nonumber\\
		&&-g_{II}{(\hat k,\hat p)}I(\bold p)-g_{IQ}{(\hat k,\hat p)}Q(\bold p)-g_{IU}{(\hat k,\hat p)}U(\bold p)-g_{IV}{(\hat k,\hat p)}V(\bold p)\bigg],
		\end{eqnarray}
		\begin{eqnarray}\label{qdotk}
		\dot{Q}(\mathbf{k})&&= \dfrac{1}{2}(\dot{\rho}_{11}-\dot{\rho}_{22}) \nonumber \\ &&=i\dot{\tau}_{_{\rm PC}}\int \frac{d\Omega}{4\pi}\bigg[f_{QI}{(\hat k,\hat p)}I(\bold p)+f_{QQ}{(\hat k,\hat p)}Q(\bold p)\nonumber\\&&-g_{QI}{(\hat k,\hat p)}I(\bold p)-g_{QQ}{(\hat k,\hat p)}Q(\bold p)-g_{QU}{(\hat k,\hat p)}U(\bold p)-g_{QV}{(\hat k,\hat p)}V(\bold p)\bigg]
		\end{eqnarray}
		\begin{eqnarray}\label{udotk}
		\dot{U}(\mathbf{k})&&=\dfrac{1}{2}(\dot{\rho}_{21}+\dot{\rho}_{12}) \nonumber \\ &&= i\dot{\tau}_{_{\rm PC}}\int \frac{d\Omega}{4\pi}\bigg[f_{UI}{(\hat k,\hat p)}I(\bold k)+f_{UU}{(\hat k,\hat p)}V(\bold k)+f_{UU}{(\hat k,\hat p)}U(\bold k)\nonumber\\
		&&-g_{UI}{(\hat k,\hat p)}I(\bold p)-g_{UQ}{(\hat k,\hat p)}Q(\bold p)-g_{UU}{(\hat k,\hat p)}U(\bold p)-g_{UV}{(\hat k,\hat p)}V(\bold p)\bigg]
		\end{eqnarray}
		\begin{eqnarray}\label{vdotk}
		\dot{V}(\mathbf{k})&&=\dfrac{1}{2}(\dot{\rho}_{21}-\dot{\rho}_{12}) \nonumber \\ &&=-i\dot{\tau}_{_{\rm PC}}\int \frac{d\Omega}{4\pi}\bigg[f_{VI}{(\hat k,\hat p)} I(\bold k)+f_{VV}{(\hat k,\hat p)} V(\bold k)\nonumber\\
		&&+g_{VI}{(\hat k,\hat p)}I(\bold p)
		+g_{VQ}{(\hat k,\hat p)}Q(\bold p)+g_{VU}{(\hat k,\hat p)}U(\bold p)+g_{VV}{(\hat k,\hat p)}Q(\bold p)+g_{VU}{(\hat k,\hat p)}U(\bold p)  \bigg] \nonumber\\
		\end{eqnarray}
		where 
		\begin{equation}\label{optical-pc10}
		\dot{\tau}_{_{\rm PC}}=\frac{3}{2} \frac{m\,v_e(\bold x) }{k^0} \sigma_T\,n_{eL}(\bold x)=\frac{3}{2} \frac{m\,v_e(\bold x) }{k^0} \sigma_T\,\delta_L n_{e}(\bold x)
		\end{equation}
		where $ v_e(\bold x)$ is electron bulk velocity and $\delta_L=\frac{n_{eL}(\mathbf{x})}{n_e(\mathbf{x})}$ is as a fraction of polarized electron number density to total one with net left-handed polarization.
		Moreover $f$'s and $g$'s are defined as
		\begin{eqnarray}
		f_{II}{(\hat{k},\hat{p})}&=& \mathcal{M}_P(1,1,1,1)+ \mathcal{M}_P(1,2,2,1)+ \mathcal{M}_P(2,1,1,2)+ \mathcal{M}_P(2,2,2,2)\nonumber\\
		f_{IQ}{(\hat{k},\hat{p})}&=& \mathcal{M}_P(1,1,1,1)+ \mathcal{M}_P(1,2,2,1)- \mathcal{M}_P(2,1,1,2)- \mathcal{M}_P(2,2,2,2)\nonumber\\
		f_{IU}{(\hat{k},\hat{p})}&=& \mathcal{M}_P(2,2,2,1)+ \mathcal{M}_P(2,1,1,1)+ \mathcal{M}_P(1,1,1,2)+ \mathcal{M}_P(1,2,2,2)\nonumber\\
		f_{IV}{(\hat{k},\hat{p})}&=& i\big(\mathcal{M}_P(2,1,1,1)+ \mathcal{M}_P(2,2,2,1)- \mathcal{M}_P(1,1,1,2)- \mathcal{M}_P(1,2,2,2)\big)\nonumber\\
		f_{QI}{(\hat{k},\hat{p})}&=& \mathcal{M}_P(1,1,1,1)+ \mathcal{M}_P(1,2,2,1)- \mathcal{M}_P(2,1,1,2)- \mathcal{M}_P(2,2,2,2)\nonumber\\
		f_{QQ}{(\hat{k},\hat{p})}&=& \mathcal{M}_P(1,1,1,1)+ \mathcal{M}_P(1,2,2,1)+ \mathcal{M}_P(2,1,1,2)+ \mathcal{M}_P(2,2,2,2)\nonumber\\
		f_{UI}{(\hat{k},\hat{p})}&=& \mathcal{M}_P(2,1,1,1)+ \mathcal{M}_P(2,2,2,1)+ \mathcal{M}_P(1,1,1,2)+ \mathcal{M}_P(1,2,2,2)\nonumber\\
		f_{UU}{(\hat{k},\hat{p})}&=& \mathcal{M}_P(2,1,1,2)+ \mathcal{M}_P(2,2,2,2)+ \mathcal{M}_P(1,1,1,1)+ \mathcal{M}_P(1,2,2,1)\nonumber\\
		f_{VI}{(\hat{k},\hat{p})}&=& i\big(\mathcal{M}_P(1,1,1,2)+\mathcal{M}_P(1,2,2,2)-\mathcal{M}_P(2,1,1,1)-\mathcal{M}_P(2,2,2,1)\big)\nonumber\\
		f_{VV}{(\hat{k},\hat{p})}&=& \mathcal{M}_P(1,1,1,1)+\mathcal{M}_P(1,2,2,1)+\mathcal{M}_P(2,1,1,2)+\mathcal{M}_P(2,2,2,2)\nonumber\\
		g_{II}{(\hat{k},\hat{p})}&=& \mathcal{M}_P(1,1,1,1)+\mathcal{M}_P(2,1,1,2)+\mathcal{M}_P(1,2,2,1)+\mathcal{M}_P(2,2,2,2)\nonumber\\
		g_{IQ}{(\hat{k},\hat{p})}&=& \mathcal{M}_P(1,1,1,1)+\mathcal{M}_P(2,1,1,2)-\mathcal{M}_P(1,2,2,1)-\mathcal{M}_P(2,2,2,2)\nonumber\\
		g_{IU}{(\hat{k},\hat{p})}&=& \mathcal{M}_P(1,2,1,1)+\mathcal{M}_P(2,2,1,2)+\mathcal{M}_P(1,1,2,1)+\mathcal{M}_P(2,1,2,2)\nonumber\\
		g_{IV}{(\hat{k},\hat{p})}&=& -i\big(\mathcal{M}_P(1,2,1,1)+\mathcal{M}_P(2,2,1,2)-\mathcal{M}_P(1,1,2,1)-\mathcal{M}_P(2,1,2,2)\big)\nonumber\\
		g_{QI}{(\hat{k},\hat{p})}&=& \mathcal{M}_P(1,1,1,1)-\mathcal{M}_P(2,1,1,2)+\mathcal{M}_P(1,2,2,1)-\mathcal{M}_P(2,2,2,2)\nonumber\\
		g_{QQ}{(\hat{k},\hat{p})}&=& \mathcal{M}_P(1,1,1,1)-\mathcal{M}_P(2,1,1,2)-\mathcal{M}_P(1,2,2,1)+\mathcal{M}_P(2,2,2,2)\nonumber\\
		g_{QU}{(\hat{k},\hat{p})}&=& \mathcal{M}_P(1,2,1,1)-\mathcal{M}_P(2,2,1,2)+\mathcal{M}_P(1,1,2,1)-\mathcal{M}_P(2,1,2,2)\nonumber\\
		g_{QV}{(\hat{k},\hat{p})}&=& i\big(\mathcal{M}_P(1,2,1,1)-\mathcal{M}_P(2,2,1,2)-\mathcal{M}_P(1,1,2,1)+\mathcal{M}_P(2,1,2,2)\big)\nonumber\\
		g_{UI}{(\hat{k},\hat{p})}&=& \mathcal{M}_P(2,1,1,1)+\mathcal{M}_P(1,1,1,2)+\mathcal{M}_P(2,2,2,1)+\mathcal{M}_P(1,2,2,2)\nonumber\\
		g_{UQ}{(\hat{k},\hat{p})}&=& \mathcal{M}_P(2,1,1,1)+\mathcal{M}_P(1,1,1,2)-\mathcal{M}_P(2,2,2,1)-\mathcal{M}_P(1,2,2,2)\nonumber\\
		g_{UU}{(\hat{k},\hat{p})}&=& \mathcal{M}_P(1,1,2,2)-\mathcal{M}_P(2,1,2,1)+\mathcal{M}_P(2,2,1,1)+\mathcal{M}_P(1,2,1,2)\nonumber\\
		g_{UV}{(\hat{k},\hat{p})}&=& i\big(\mathcal{M}_P(1,1,2,2)-\mathcal{M}_P(2,1,2,1)-\mathcal{M}_P(2,2,1,1)-\mathcal{M}_P(1,2,1,2)\big)\nonumber\\
		g_{VI}{(\hat{k},\hat{p})}&=& i\big(\mathcal{M}_P(1,1,1,2)-\mathcal{M}_P(2,1,1,1)+\mathcal{M}_P(1,2,2,2)-\mathcal{M}_P(2,2,2,1)\big)\nonumber\\
		g_{VQ}{(\hat{k},\hat{p})}&=& i\big(\mathcal{M}_P(1,1,1,2)-\mathcal{M}_P(2,1,1,1)-\mathcal{M}_P(1,2,2,2)+\mathcal{M}_P(2,2,2,1)\big)\nonumber\\
		g_{VU}{(\hat{k},\hat{p})}&=& i\big(\mathcal{M}_P(1,2,1,2)-\mathcal{M}_P(2,2,1,1)+\mathcal{M}_P(1,1,2,2)-\mathcal{M}_P(2,1,2,1)\big)\nonumber\\
		g_{VV}{(\hat{k},\hat{p})}&=& \mathcal{M}_P(1,2,1,2)-\mathcal{M}_P(2,2,1,1)-\mathcal{M}_P(1,1,2,2)+\mathcal{M}_P(2,1,2,1)
		\end{eqnarray}
		where we bring some of the above functions after tedious but straightforward calculations as follows
		\begin{eqnarray}
		f_{IV}{(\hat{k},\hat{p})} &=& \hat{\bold v}.\epsilon_1(\bold k)\Big(\hat {\bold q}.\epsilon_2(\bold k)\times\epsilon_1(\bold p)\hat{\bold k}.\epsilon_1(\bold p)+\hat{\bold v}.\epsilon_2(\bold k)\times\epsilon_2(\bold p)\hat{\bold k}.\epsilon_2(\bold p)-\hat{\bold v}.\hat{\bold k}\times\epsilon_1(\bold p)\epsilon_2(\bold k).\epsilon_1(\bold p)\nonumber\\
		&-&\hat{\bold v}.\hat{\bold k}\times\epsilon_2(\bold p)\epsilon_2(\bold k).\epsilon_2(\bold p)+\hat{\bold v}.\epsilon_1(\bold k)\Big)-(\hat{\bold v}.\epsilon_1(\bold p))^2-(\hat{\bold v}.\epsilon_2(\bold p))^2\nonumber\\
		&+&\hat{\bold v}.\epsilon_2(\bold k)\Big(\hat{\bold v}.\epsilon_1(\bold p)\epsilon_2(\bold k).\epsilon_1(\bold p)+\hat{\bold v}.\epsilon_2(\bold p)\epsilon_2(\bold k).\epsilon_2(\bold p)\Big)\nonumber\\
		&+&\hat{\bold p}.\epsilon_2(\bold k)\Big(\hat{\bold v}.\epsilon_1(\bold p)\hat{\bold v}.\epsilon_1(\bold k)\times\epsilon_1(\bold p)+\hat{\bold v}.\epsilon_2(\bold p)\hat{\bold v}.\epsilon_1(\bold k)\times\epsilon_2(\bold p)\Big)\nonumber\\
		&+& \hat {\bold v}.\epsilon_2(\bold k)\Big(\hat{\bold v}.\epsilon_1(\bold p)\epsilon_1(\bold k).\epsilon_2(\bold p)-\hat{\bold v}.\epsilon_2(\bold p)\epsilon_1(\bold k).\epsilon_1(\bold p)\Big)\nonumber\\
		&+&\hat{\bold v}.\hat{\bold p}\times\epsilon_1(\bold k)\Big(\hat{\bold v}.\epsilon_1(\bold p)\epsilon_2(\bold k).\epsilon_1(\bold p)+\hat{\bold v}.\epsilon_1(\bold p)\epsilon_2(\bold k).\epsilon_1(\bold p)\Big)\nonumber\\
		&-&\bigg(\hat{\bold v}.\epsilon_2(\bold k)\Big(\hat{\bold v}.\epsilon_1(\bold k)\times\epsilon_1(\bold p)\hat{\bold k}.\epsilon_1(\bold p)+\hat{\bold v}.\epsilon_1(\bold k)\times\epsilon_2(\bold p)\hat{\bold k}.\epsilon_2(\bold p)-\hat{\bold v}.\hat{\bold k}\times\epsilon_1(\bold p)\epsilon_1(\bold k).\epsilon_1(\bold p)\nonumber\\
		&-&\hat{\bold v}.\hat{\bold k}\times\epsilon_2(\bold p)\epsilon_1(\bold k).\epsilon_2(\bold p)-\hat{\bold v}.\epsilon_1(\bold k)\Big)+(\hat{\bold v}.\epsilon_1(\bold p))^2+(\hat{\bold v}.\epsilon_2(\bold p))^2\nonumber\\
		&-&\hat{\bold v}.\epsilon_1(\bold k)\Big(\hat{\bold v}.\epsilon_1(\bold p)\epsilon_1(\bold k).\epsilon_1(\bold p)+\hat{\bold v}.\epsilon_2(\bold p)\epsilon_1(\bold k).\epsilon_2(\bold p)\Big)\nonumber\\
		&+&\hat{\bold p}.\epsilon_1(\bold k)\Big(\hat{\bold v}.\epsilon_1(\bold p)\hat{\bold v}.\epsilon_2(\bold k)\times\epsilon_1(\bold p)+\hat{\bold v}.\epsilon_2(\bold p)\hat{\bold v}.\epsilon_2(\bold k)\times\epsilon_2(\bold p)\Big)\nonumber\\
		&+&\hat{\bold v}.\epsilon_1(\bold k)\Big(\hat{\bold v}.\epsilon_1(\bold p)\epsilon_2(\bold k).\epsilon_2(\bold p)-\hat{\bold v}.\epsilon_2(\bold p)\epsilon_2(\bold k).\epsilon_1(\bold p)\Big)\nonumber\\
		&+&\hat{\bold v}.\hat{\bold p}\times\epsilon_2(\bold k)\Big(\hat{\bold v}.\epsilon_1(\bold p)\epsilon_1(\bold k).\epsilon_1(\bold p)+\hat{\bold v}.\epsilon_2(\bold p)\epsilon_1(\bold k).\epsilon_1(\bold p)\Big)\bigg) \\
		f_{UV}{(\hat{k},\hat{p})} &=& \hat{\bold v}.\epsilon_1(\bold k)\Big(\hat{\bold v}.\epsilon_1(\bold k)\times\epsilon_1(\bold p)\hat{\bold k}.\epsilon_1(\bold p)+\hat{\bold v}.\epsilon_1(\bold k)\times\epsilon_2(\bold p)\hat{\bold k}.\epsilon_2(\bold p)\nonumber\\
		&-&3\Big(\hat{\bold v}.\epsilon_1(\bold p)\epsilon_1(\bold p).\epsilon_2(\bold k)+\hat{\bold v}.\epsilon_2(\bold p)\epsilon_2(\bold p).\epsilon_2(\bold k)\Big)\nonumber\\
		&-&\hat{\bold v}.\hat{\bold k}\times\epsilon_1(\bold p)\epsilon_1(\bold k).\epsilon_1(\bold p)
		-\hat{\bold v}.\hat{\bold k}\times\epsilon_2(\bold p)\epsilon_1(\bold k).\epsilon_2(\bold p)\Bigg)\nonumber\\
		&+&\bold q.\epsilon_2(\bold k)\Bigg(\hat{\bold v}.\epsilon_2(\bold k)\times\epsilon_1(\bold p)\bold k.\epsilon_1(\bold p)+\hat{\bold v}.\epsilon_2(\bold k)\times\epsilon_2(\bold p)\bold k.\epsilon_2(\bold p)\nonumber\\
		&+&3\Big(\hat{\bold v}.\epsilon_1(\bold p)\epsilon_1(\bold p).\epsilon_1(\bold k)+\hat{\bold v}.\epsilon_2(\bold p)\epsilon_2(\bold p).\epsilon_1(\bold k)\Big)\nonumber\\
		&-&\hat{\bold v}.\hat{\bold k}\times\epsilon_1(\bold p)\epsilon_2(\bold k).\epsilon_1(\bold p)
		-\hat{\bold v}.\hat{\bold k}\times\epsilon_2(\bold p)\epsilon_2(\bold k).\epsilon_2(\bold p)\Bigg)\nonumber\\
		&+&\hat{\bold v}.\epsilon_1(\bold k)\Big(\hat{\bold v}.\epsilon_1(\bold p)\epsilon_1(\bold k).\epsilon_2(\bold p)-\hat{\bold v}.\epsilon_2(\bold p)\epsilon_1(\bold k).\epsilon_1(\bold p)\Big)\nonumber\\
		&+&\hat{\bold v}.\epsilon_2(\bold k)\Big(\hat{\bold v}.\epsilon_1(\bold p)\epsilon_2(\bold p).\epsilon_2(\bold k)
		-\hat{\bold v}.\epsilon_2(\bold p)\epsilon_2(\bold k).\epsilon_1(\bold p)\Big)\nonumber\\
		&+&\hat{\bold p}.\epsilon_1(\bold k)\Big(\hat{\bold v}.\epsilon_1(\bold p)\hat{\bold v}.\epsilon_1(\bold k)\times\epsilon_1(\bold p)+\hat{\bold v}.\epsilon_2(\bold p)\hat{\bold v}.\epsilon_1(\bold k)\times\epsilon_2(\bold p)\Big)\nonumber\\
		&+&\hat{\bold p}.\epsilon_2(\bold k)\Big(\hat{\bold v}.\epsilon_1(\bold p)\hat{\bold v}.\epsilon_2(\bold k)\times\epsilon_1(\bold p)+\hat{\bold v}.\epsilon_2(\bold p)\hat{\bold v}.\epsilon_2(\bold k)\times\epsilon_2(\bold p)\Big)\nonumber\\
		&+&\hat{\bold v}.\hat{\bold p}\times\epsilon_1(\bold k)\Big(\hat{\bold v}.\epsilon_1(\bold p)\epsilon_1(\bold k).\epsilon_1(\bold p)+\hat{\bold v}.\epsilon_2(\bold p)\epsilon_1(\bold k).\epsilon_2(\bold p)\Big)\nonumber\\
		&+&\hat{\bold v}.\hat{\bold p}\times\epsilon_2(\bold k)\Big(\hat{\bold v}.\epsilon_1(\bold p)\epsilon_2(\bold k).\epsilon_1(\bold p)+\hat{\bold v}.\epsilon_2(\bold p)\epsilon_2(\bold k).\epsilon_2(\bold p)\Big)
		\end{eqnarray}
		Since other functions have too long formulation, we neglected to write all of them here. But, we consider them to derive the B-mode and E-mode power spectrum of CMB.
	\end{document}